\documentclass[pra,showpacs,twocolumn]{revtex4}
\usepackage{bm}
\usepackage{amsmath}
\usepackage{amssymb}
\usepackage{graphicx}

\newcommand{\eqb}{\begin{eqnarray}}
\newcommand{\eqe}{\end{eqnarray}}
\newcommand{\rd}{\mathrm{d}}
\newcommand{\bx}{{\bf x}}
\newcommand{\bk}{{\bf k}}
\newcommand{\bq}{{\bf q}}
\newcommand{\br}{{\bf r}}
\newcommand{\bQ}{{\bf Q}}
\newcommand{\rX}{{\rm X}}

\newcommand{\pd}{\partial}

\begin{document}

\title{Nonlinear tunneling of BEC in an optical lattice: signatures of quantum collapse and revival}
\author{ V. S. Shchesnovich$^1$ and V. V. Konotop$^{2}$ }
\affiliation{
$^1$ Instituto de F\'{\i}sica - Universidade Federal de Alagoas, Macei\'o AL
57072-970, Brazil\\
$^2$Centro de F\'isica Te\'orica e Computacional, Universidade de Lisboa, Complexo
Interdisciplinar, Avenida Professor Gama Pinto 2, Lisboa 1649-003, Portugal;
Departamento de F\'isica, Faculdade de Ci\^encias, Universidade de Lisboa, Campo
Grande, Ed. C8, Piso 6, Lisboa 1749-016, Portugal, and Departamento de
Matem\'aticas, E. T. S. de Ingenieros Industriales, Universidad de Castilla-La
Mancha 13071 Ciudad Real, Spain }

\begin{abstract}
Quantum theory of the intra-band resonant tunneling  of a Bose-Einstein  condensate
loaded in a two-dimensional optical lattice is considered. It is shown that the
phenomena of quantum collapse and revival can be observed in the fully quantum
problem. The meanfield limit of the theory is analyzed using the WKB approximation
for discrete equations, establishing in this way a direct connection between the
two approaches conventionally used in very different physical contexts. More
specifically we show that there exist two different regimes of tunneling and study
dependence of quantum collapse and revival on the number of condensed atoms.
\end{abstract}

\maketitle

\section{Introduction}

Passage from the exact many-body description of an atomic gas at zero temperature
to the mean-field theory is based on the assumption about large occupation number
$N$ of the ground (or, more generally, some specific quantum) state. Then
$N^{-1/2}$ can be shown to be the parameter of expansion, which in the leading
order gives the Gross-Pitaevskii equation~\cite{Gardiner}. Respectively, the
mentioned requirement must be also applied to mean-field theories of the
Landau-Zener tunneling~\cite{Landau-Zener,KKS,SHCH}, for which interesting
experimental data are  available~\cite{experiment}, to instabilities of BEC's in
optical lattices~\cite{KKS,instabil} observed experimentally in
\cite{instb-experim}, and to a theory of nonlinear Bloch--band
tunneling~\cite{BKK}. Similar approach was also exploited in a pure quantum theory
of a Bose-Einstein condensate (BEC) in a double-well trap~\cite{JM},  based on a
model with linear coupling~\cite{JZM}. Either for the Bloch--band tunneling problem
or for the tunneling in a double-well potential the approach of large $N$ is
justified when the system is close to the mean-field regime and the respective
dynamical solutions are characterized by the essentially nonzero populations of the
quantum states between which tunneling occurs. Such regimes indeed exist when the
description is reduced to a dimer~\cite{RSFS}. At the same  time the mean-field
approximation to the above problems allows solutions where the populations of the
states repeatedly become zero. Strictly speaking this violates the initially made
supposition about large atomic numbers, because a state with negligible population
cannot be treated in the mean-filed approximation. Therefore the theory requires
modifications.

We thus can formulate the main goal of  the present paper as the analysis of the
macroscopic nonlinear  tunneling between two quantum states with the same (or
infinitely close) energies in the limit corresponding to large  total number of
particles, allowing however populations of each of the states to become negligibly small
at some moments of time. We will pay the main attention to the quantum effects, not
accounted by the standard mean-field theory.

\begin{figure}[ht]
\begin{center}
\includegraphics[scale=0.5,  bb = 118   272   405   557 ]{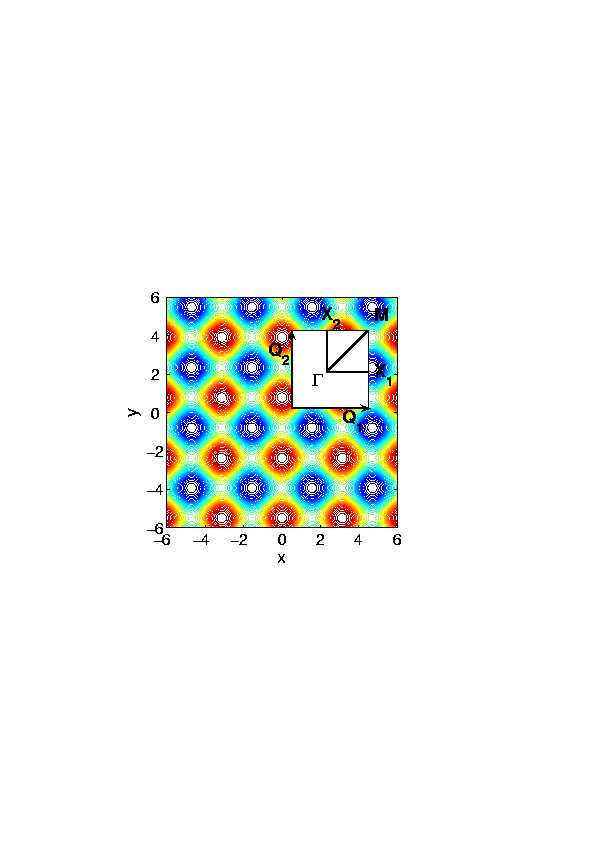}
\caption{(Color online) The contour plot of the
periodic potential, given by equation (\ref{LATT}) below, used for the numerical
simulations. The inset shows the first Brillouin zone, the reciprocal lattice basis
vectors ${\bf Q}_{1,2}$ and the high symmetry points $\Gamma$, M, and X${}_j$. The
nonlinear intra-band tunneling occurs between the points X$_1$ and X$_2$.}
\label{FG0}
\end{center}
\end{figure}

To be specific, as a physical system  we explore a BEC loaded in a two-dimensional
optical lattice and consider tunneling between two states with the same energy.
Such a situation can be experimentally realized in at least two different ways.
First, one can use a non-separable lattice, with properly chosen parameters
providing closer of the lowest gap (as, for example, this is suggested
in~\cite{BKK}). Then nonlinear tunneling occurs between X and M points of two
different bands (see Fig.~\ref{FG0}). More general situation, however, corresponds
to the nonlinear tunneling between two X points of the same band, rotated by
$\pi/2$ with respect to each other (these are the points X$_{1,2}$ in
Fig.~\ref{FG0}). Such intra--band  tunneling does not depend on the size of the
gap, and hence can be observed in any lattice including separable one. Moreover,
whenever one deals with the nonlinear tunneling, the stability issue acquires
especial importance~\cite{KKS,BKK}, since depending on the sign of the effective
mass the homogeneous excitations can be either modulationally stable or unstable.
This in particular means that tunneling between two states in different bands (like
tunneling between X and M points, mentioned above, or,  tunneling in
one-dimensional case between two neighbor zones) represents a transition between
stable and unstable states, what causes asymmetry of the tunneling. That is why, in
the present paper we concentrate on intra--band tunneling between two stable X
points, what rules out instability and allows one to limit the consideration to
plane Bloch waves.

In the mean-field approximation this problem was  considered in~\cite{BKKS}, which
in the two-mode approximation is reduced to a model imitating either a Josephson
junction or Rabi oscillations in a system of two-level atoms. Such a
model describes population oscillations between two states.  Using the mentioned
analogy, one can predict that quantization of the motion should result in quantum
collapse and revival of the above oscillations~\cite{CBPZ}. Obtaining those
phenomena in the process of intra-band tunneling, constitutes  the second aim of the present
paper.

Finally, we  emphasize an interesting mathematical  aspect of the problem at hand.
Namely, we will establish a formal link between the mean-field approximation, where
the small parameter can be identified as $1/N$ and the discrete WKB method, having
$\hbar$ as a formal small parameter, which is described in details in
Ref.~\cite{braun}. In our approach the both methods, appearing originally as
independent as they are used in different setups and based on different physical
assumptions, are intimately related  to each other allowing us to appreciate the
accuracy of the mean-field model, what will be the third goal of the present work.
It is to be mentioned that earlier the link  between the semi-classical and
mean-field approaches was discussed in~\cite{VA}, where a linearly couped two-level
system, modeling either a BEC in a double-well trap or a spinor condensate, was
considered and where the quantum corrections appeared as a decoherence of the
quantum states. We, thus present one more physical system, a BEC in an optical
lattice, having no linear coupling and rather different stability properties, which
also allows experimental verification of the mean-field approximation. In our case
the quantum effects will be manifested through the quantum collapse.

The paper is organized as follows. In Sec. II we  deduce the quantum two-mode model
describing resonant intra-band tunneling, expand its solution over the Fock basis
and deduce the dynamical system for the expansion coefficients. In Sec. III we
discuss the mean-field approximation starting with the derived dynamical system and
based on the WKB method for discrete equations. Sec. IV is devoted to numerical
study of the both, full quantum and mean-field models. The outcomes are summarized
in the Conclusion.

\section{Quantum model}

\subsection{Inter-- and intra-- band transitions}

Let us start with the Hamiltonian of a BEC in a two-dimensional, ${\bf x}\in\mathbb{R}^2$, optical lattice $V(\bx)$,
\eqb
H = \int\limits_\mathcal{V}\rd^2\bx
\psi^\dag(\bx)\left(-\frac{\hbar^2}{2m}\nabla^2+V(\bx)\right)\psi(\bx)
\nonumber \\
+\frac{g}{2}\int\limits_\mathcal{V}\rd^2\bx
\psi^\dag(\bx)\psi^\dag(\bx)\psi(\bx)\psi(\bx),
\label{EQ1}
\eqe
where $\mathcal{V}=Mv_0$ is the total area of the lattice consisting of $M$ cells each one
of the area $v_0$, $g$ is the  interaction coefficient in two dimensions, and
$\psi^\dag({\bf x})$ and $\psi({\bf x})$ are the creation and annihilation field operators.
Introducing the Bloch waves
$\varphi_{n\bk}(\bx)
$ through the standard eigenvalue problem
$\left(-\frac{\hbar^2}{2m}\nabla^2+V(\bx)\right)\varphi_{n\bk}=E_{n\bk}\varphi_{n\bk}$,
where $n$ is a number of the zone and $\bk$ is the wave-vector in the first
Brillouin zone (BZ), and considering them orthonormalized,
$\int_{\mathcal{V}}\bar{\varphi}_{n\bk}({\bf x})\varphi_{n'\bk'}({\bf x})d{\bf
x}=\delta_{nn'}\delta_{\bk\bk'}$, we represent
\eqb
\psi(\bx) = \sum_{n,\bk}\varphi_{n,\bk}(\bx)b_{n,\bk},
\label{EQ2}
\eqe
where the creation and annihilation operators satisfy the usual commutation relations
$[b_{n\bk},b_{n^\prime\bk^\prime}]=0$ and $[b_{n\bk},b^\dag_{n^\prime\bk^\prime}] =
\delta_{nn^\prime}\delta_{\bk\bk^\prime}$.

The expansion (\ref{EQ2}) allows one to rewrite the Hamiltonian in the form
\eqb
&&H = \sum_{n,\bk}E_{n\bk}b^\dag_{n\bk} b_{n\bk} +
\nonumber \\
&&
\sum_{\bk_1,...,\bk_4 \atop n_1,...,n_4}
\chi^{n_1n_2n_3n_4}_{\bk_1\bk_2\bk_3\bk_4}
\delta_{\bk_1+\bk_2-\bk_3-\bk_4,\bQ}b^\dag_{n_1\bk_1} b^\dag_{n_2\bk_2}
b_{n_3\bk_3}b_{n_4\bk_4},
\nonumber \\
\label{EQ3}\eqe
where $\bQ$ is an arbitrary vector of the reciprocal lattice,
\eqb
\chi^{n_1n_2n_3n_4}_{\bk_1\bk_2\bk_3\bk_4} = \frac{g}{2}
\int\limits_{\mathcal{V}}\rd^2\bx
\varphi^*_{n_1\bk_1}\varphi^*_{n_2\bk_2}\varphi_{n_3\bk_3}\varphi_{n_4\bk_4}
\label{EQ4}
\eqe
(hereafter an asterisk stands for complex conjugation).

Next we apply the rotating wave approximation, i.e.  neglect all
four-wave processes which do not satisfy the energy conservation law.
This allows us to simplify the Hamiltonian (\ref{EQ3}):
\eqb
&&H = \sum_{n,\bk}E_{n\bk}b^\dag_{n\bk} b_{n\bk} +
\nonumber \\
&& \sum_{\bk_1,...,\bk_4 \atop n_1,...,n_4}
\delta_{\bk_1+\bk_2,\bk_3+\bk_4+\bQ}\delta_{E_{n_1\bk_1}+E_{n_2\bk_2},E_{n_3\bk_3}+E_{n_4\bk_4}}
\nonumber \\
&& \times\chi^{n_1n_2n_3n_4}_{\bk_1\bk_2\bk_3\bk_4}
b^\dag_{n_1\bk_1} b^\dag_{n_2\bk_2}
b_{n_3\bk_3}b_{n_4\bk_4}.
\label{EQ3_1}
\eqe

Further reduction of the Hamiltonian can be done by taking into account that the
periodic medium is highly dispersive, and therefore for the {\em resonant}
four-wave interactions, equality of the group velocities of the respective matter
waves must be imposed.  In the case of tunneling between two states, say
$(n_1,\bq_1)$ and $(n_2,\bq_2)$, this means the constraint $\partial
E_{n_1\bq_1}/\partial \bq_1=\partial E_{n_2\bq_2}/\partial \bq_2$. Together with
the momentum and energy conservation laws, expressed by the first and the second
Kronecker deltas in (\ref{EQ3_1}), this means that the resonant nonlinear tunneling
can occur only between highly symmetric points of the BZ. As far as we have limited
ourselves to the intra-band tunneling, we conclude that the only points of the BZ
-- the points X -- satisfies the above requirements. Moreover, in the BZ there are
only two physically different X-points. Assuming that the lattice is orthogonal
with a period $d$ in the both directions, the BZ can be identified simply as the
domain $[-\frac{\pi}{d},\frac{\pi}{d}]\times [-\frac{\pi}{d},\frac{\pi}{d}]$.
Consequently  the two X-points we are interested in, below they are referred to as
X$_1$ and X$_2$,    correspond to the two vectors
$\bq_1=\left(0,\frac{\pi}{d}\right)$ and $\bq_2=\left(\frac{\pi}{d},0\right)$. The
group velocities in these points are zero.

Now one readily verifies that (\ref{EQ3_1}) is significantly simplified
allowing one to rewrite the Hamiltonian in the form of a sum
\begin{eqnarray}
\label{H1}
    H= \sum_nH_n+H_{i-b}
\end{eqnarray}
where
\begin{widetext}
\begin{eqnarray}
H_n=E_{n\bq_1}b^\dag_{n\bq_1} b_{n\bq_1} +E_{n\bq_2}b^\dag_{n\bq_2} b_{n\bq_2} + \chi_{\bq_1\bq_1}^{nn}\left(b^\dag_{n\bq_1} b^\dag_{n\bq_1}
b_{n\bq_1}b_{n\bq_1}+b^\dag_{n\bq_2} b^\dag_{n\bq_2}
b_{n\bq_2}b_{n\bq_2}\right)
\nonumber \\
+\chi_{\bq_1\bq_2}^{nn}\left(b^\dag_{n\bq_1} b^\dag_{n\bq_1}
b_{n\bq_2}b_{n\bq_2}+b^\dag_{n\bq_2} b^\dag_{n\bq_2}
b_{n\bq_1}b_{n\bq_1}+4b^\dag_{n\bq_1} b^\dag_{n\bq_2}
b_{n\bq_1}b_{n\bq_2}\right)
\end{eqnarray}
describes intra-band transitions between states X$_1$ and X$_2$ of the $n$-th band and
\begin{eqnarray}
H_{i-b}= 4\sum_{n_1n_2\atop n_1\neq n_2}
\chi^{n_1n_2}_{\bq_1\bq_2}\left(b^\dag_{n_1\bq_1}
b^\dag_{n_2\bq_2}+b^\dag_{n_2\bq_1} b^\dag_{n_1\bq_2}\right)
\left(b_{n_1\bq_1}b_{n_2\bq_2}+b_{n_2\bq_1}b_{n_1\bq_2}\right)
\end{eqnarray}
\end{widetext}
describes transitions between X points of different bands.  In the above formulas
we introduced $\chi^{n_1n_2}_{\bq_1\bq_2}
=\chi_{\bk_1\bk_2\bk_1\bk_2}^{n_1n_2n_1n_2}$ and used the symmetry properties of
the Bloch functions,  giving in particular
$\chi_{\bq_1\bq_1}^{nn}=\chi_{\bq_2\bq_2}^{nn}$ and
$\chi_{\bq_1\bq_2}^{nn}=\chi_{\bq_2\bq_1}^{nn}$, as well as the fact that
$\varphi_{n\bq_1}$ and $\varphi_{n\bq_2}$ correspond to X points and hence can be
chosen real and periodic.

\subsection{Dynamical equations and their accuracy}

Dynamics of the condensate can be described by the
time evolution of the coefficients of the expansion of a given multiparticle state $|\Psi
(t)\rangle$ over the Fock basis. In the case when initially one band, say the band
$n_0$, is populated more densely than the other ones, it is natural to represent the
basis as $|N_1,N_2;{\bf n}\rangle$, where $N_j$ stand for the occupation numbers of
the states $\rX_j$ of the given band, while ${\bf n}$ symbolically designate
occupation numbers of all other bands. Thus
\begin{eqnarray}
\label{psi_t_1}
    |\Psi (t)\rangle=\sum_{N_1,N_2,{\bf n}}C_{N_1N_2{\bf n}}(t)|N_1,N_2;{\bf n}\rangle
\end{eqnarray}

Now we consider the situation, where initially  (i.e. at $t=0$) all atoms
belong to X points of the chosen band, i.e. when $N_1+N_2=N$ with $N$ being the total number of
particles. Then, introducing the simplified notation $|N_1,N_2\rangle=|N_1,N_2;{\bf 0}\rangle$ for such
states, which also can be rewritten as $|N_1,N_2\rangle=|k,N-k\rangle=
|k\rangle|N-k\rangle$ (here $k=N_1$), and
respectively $C_{N_1N_20}(t)\equiv C_k(t)$,  we can express
\begin{eqnarray}
\label{psi-init}
    |\Psi (0)\rangle=\sum_{k=0}^NC_{k}(0)|k,N-k\rangle, \quad \sum_{k=0}^N|C_k(0)|^2 = 1.
\end{eqnarray}

For the next step, it is not difficult to verify  that the dynamics  of a state,
initially spanned over the kets $|k,N-k\rangle$, i.e. subject to the initial
condition (\ref{psi-init}), which is induced by the Hamiltonian (\ref{H1}) results
only in the states belonging to initial sub-space, i.e.
\begin{eqnarray}
\label{psi-time}
    |\Psi (t)\rangle=\sum_{k=0}^NC_{k}(t)|k,N-k\rangle, \quad \sum_{k=0}^N|C_k(t)|^2 = 1.
\end{eqnarray}
for all $t$.

Indeed, substituting (\ref{psi_t_1}) in the  Shr\"odinger equation
\begin{eqnarray}
\label{SE}
i\frac{\partial |\Psi\rangle}{\partial t}=H|\Psi\rangle
\end{eqnarray}
 and applying
$\langle N_1,N_2;{\bf n}|$ in order to obtain the equations for the expansion
coefficients, one finds that for all coefficients with $N_1+N_2<N$ (i.e. for states
with ${\bf n}\neq 0$) the derivatives $dC_{N_1N_2{\bf n}}/dt$ linearly depend on
different $C_{N_1'N_2'{\bf n}'}$ with $N_1'+N_2'<N$, but do not depend  on
$C_{N_1N_20}$. This  follows from the relations $H_{i-b}|N_1,N_2\rangle=0$  and
$H_n|N_1,N_2\rangle=0$ for $n\neq n_0$. The first of these formulas means that no
coupling occurs when at least one of the states in not occupied, while the second
relation means zero result when one probes the energy of an empty band ($H_n$ does
not originate inter-band transitions) and in our case the only band, $n_0=0$, is
initially occupied. Thus all the  coefficients $C_{N_1N_2{\bf n}}$ with ${\bf
n}\neq 0$ are identically equal to zero, provided they are zero at $t=0$.

Summarizing the described situation subject to assumption that  only one band $n_0$
is populated and neglecting the unessential for the dynamics linear energy term
$E_{n_0\bq_1}+E_{n_0\bq_2}$, one arrives at the two-mode model whose Hamiltonian
can be written down in the form
\begin{eqnarray}
&&H_{n_0}= H_0 -E_{n_0\bq_1}-E_{n_0\bq_2}
\nonumber \\
&&=\chi_{11}\left\{n_1^2 +n_2^2 + \Lambda\left[4n_1n_2 +
(b_1^\dag)^2b_2^2 +(b_2^\dag)^2b_1^2\right]\right\},
\nonumber\\
& &
\label{EQ7}
\end{eqnarray}
where $b_j=b_{n_0\bq_j}$, i.e. $b^\dag_1|k,N-k\rangle = \sqrt{k+1}|k+1,N-k\rangle$
and $b^\dag_2|k,N-k\rangle = \sqrt{N-k+1}|k,N-k+1\rangle$, $n_j = b^\dag_{n_0\bq_j}
b_{n_0\bq_j}$  are the populations of the states $X_j$ of the band $n_0$, $\chi_{11}=\chi_{q_1q_1}^{n_0n_0}$,  and
$\Lambda = \chi_{\bq_1\bq_2}^{n_0n_0}/\chi_{\bq_1\bq_1}^{n_0n_0}$. Hamiltonian
(\ref{EQ7}) preserves the total number of atoms in the $X_j$-points, what naturally
reflects the approximations made.

The above calculations, which resulted in (\ref{EQ7}),  hold for the more general
Hamiltonian (\ref{EQ3_1}) obtained in the rotating wave approximation, but fail for
the original model accounting all possible transitions (not only the resonant ones). This defines  the accuracy of the two-mode
approximation: the ratios  between the accounted and neglected  terms are of the
order of $1/N$.

Now the Schr\"odinger equation (\ref{SE}) with $H=H_0$ results in a system  of
ordinary differential equations for the coefficients $C_k$
\eqb
\frac{i}{N}\frac{\rd C_k}{\rd \tau} =
\frac{\Lambda}{4}(b_{k-1}C_{k-2}+b_{k+1}C_{k+2}) + \frac{a_k}{4}C_k,
\label{EQ9}\eqe
\medskip
with the dimensionless time $\tau =(4\chi_{11} N/\hbar)t$ and with the coefficients
\eqb
&&a_k =1+  2(2\Lambda-1)
\frac{k}{N}\left(1-\frac{k}{N}\right),
\\
&&b_k =
\left[\frac{k}{N}\left(\frac{k}{N}+\frac{1}{N}\right)
\left(1-\frac{k}{N}\right)\left(1-\frac{k}{N}+\frac{1}{N}\right)\right]^{\frac{1}{2}}.
\label{EQ10}
\eqe

\section{Fictitious particle representation and the semiclassical limit}
\label{semiclass}

\subsection{Mean-field equations viewed as a quasi-classical limit.}

Equation (\ref{EQ9}) can be viewed as the Schr\"odinger equation for a fictitious
quantum particle in the one-dimensional discrete space $k = \{0,1,2,\ldots,N\}$.
Indeed, setting $x = k/N$ and $h = 2/N$, introducing a differentiable  function
$\psi(x)$ of the continuous variable $x\in(0,1)$, such that $\psi(x) = \sqrt{N}C_k$
at the points $x=k/N$~\footnote{One can always draw a smooth curve through a finite
number $N$ of points, say, by a $(N-1)$-th order polynomial. However, this imposes
a constraint on the initial state $\psi(x,0)$. On the other hand, for  a non-smooth
initial state $\psi(x,0)$ the limit $h\to 0$ may  either do not lead to the classical dynamics or do not exist.}
let us define the quantities:
\begin{eqnarray*}
&&b^{(\pm)}_h(x) = b_h(x+h/2)\pm b_h(x-h/2),
\\
&&b_h(x) = [x(x+h/2)(1-x)(1-x+h/2)]^{1/2} ,
\\
&&a(x) = 1 + 2(2\Lambda-1)b_0(x).
\end{eqnarray*}
Now equation (\ref{EQ9}) becomes
\eqb
&&ih\pd_\tau \psi(x) =\frac{a(x)}{2}\psi(x)
\nonumber \\
&&+ \frac{\Lambda}{2}\left\{b_h\left(x+\frac h2\right)e^{i\hat{p}} +
b_h\left(x-\frac h2\right)e^{-i\hat{p}}\right\}\psi(x).\nonumber \\
&& \label{EQ11}
\eqe
Here the parameter $h$ plays the role of the Plank constant $\hbar$,  $\hat{p}
\equiv -ih\pd_x$ is the ``momentum operator'' of the fictitious particle, and the
``wave function'' $\psi(x)$ is assumed to be normalized in the usual way
$\int_0^1\rd x |\psi(x)|^2=1$.

Equation (\ref{EQ11}) describes a fictitious quantum particle of the mass
proportional to $1/\Lambda$ moving in a compact curved space defined by the
interval $(0,1)$ (hence $b^{(\pm)}_h(x)$ at the functions of the  momentum
operator). Note that the ``momentum'' eigenvalues can be restricted to the first
BZ, i.e.  $[-\frac{\pi}{d},\frac{\pi}{d}]$.

The semiclassical dynamics corresponds to the limit $h\to 0$, i.e. when the number
of BEC atoms $N\to\infty$, what is the usual limit of the Gross-Pitaevskii
equation. It should be noted that  the characteristic time $t$ of the evolution
scales as $(gN)^{-1}\tau$, hence the quantity $gN$ must be kept fixed, which is the
second condition of the mean-field limit. It is also clear that the limit $h\to0$,
if it exists, corresponds to the continuous limit of the discrete equation
(\ref{EQ11}) (see also Ref. \cite{braun}).

In order to derive the quasi-classical equation corresponding to the limit $h\to0$
we proceed in the usual way. Setting $\psi(x,\tau) = e^{iS(x,\tau,h)/h}$  for a
complex action $S(x,\tau,h)$ viewed as a series $S = S^{(0)} +hS^{(1)} + \ldots$,
we get the equation
\begin{widetext}
\eqb
-S_\tau
= \frac{\Lambda}{2}\left\{b_h\left(x+\frac h2\right)e^{i\frac{S(x+h)-S(x)}{h}} +
b_h\left(x-\frac h2\right)e^{i\frac{S(x-h)-S(x)}{h}}\right\} + \frac{a(x)}{2}.
\label{EQ12}\eqe
\end{widetext}
Assuming that the function $S(x,\tau,h)$ has derivatives with respect to $x$ up to
the second order, expanding
\[
\frac{S(x\pm h)- S(x)}{h} = \pm S^{(0)}_x + \mathcal{O}(h)
\]
and setting $h\to0$ in the equation (\ref{EQ12}) we get the Hamilton-Jacobi
equation for the classical action $\tilde{S}(x,\tau)=S^{(0)}(x,\tau)-\tau/2$:\
\eqb
-\tilde{S}_\tau = b_0(x)\left\{\Lambda\cos(\tilde{S}_x) +2\Lambda -1\right\}
\label{EQ13}\eqe
[recall that $b_0(x) = x(1-x)$].

Thus the physical sense of the transition to the classical limit in our problem is
the transition from a discrete equation  to its continuous limit. Note that in our
setup the usual classical limit $\hbar\to0$, sometimes understood as a mathematical
abstraction, acquires the well established sense of the limit of a large number of atoms  and thus
{\em can be studied experimentally}.

In the case of a finite $h$ (finite $N$) we have $[\hat{p},x] = -ih$, i.e. the
usual canonical commutator of the momentum and coordinate. It turns out, however,
that quasi-classical dynamics is more convenient to describe in terms of variables
$z = 1-2 x $ and $\Phi = \tilde{S}_x=p$, where $x$ and $p$ are the classical limits
of the corresponding quantum variables. Then the Poisson brackets of the respective
classical dynamics read
$$
\{\Phi,z\} = \lim_{h\to0}\frac{i}{h}[\hat{p},1-2x] = -2.
$$
and the classical Hamiltonian can be recovered
from the Hamilton-Jacobi equation (\ref{EQ13}):
\eqb
\mathcal{H} = \frac{1}{4}(1-z^2)(2\Lambda +\Lambda\cos\Phi - 1).
\label{EQ14}\eqe
Thus the classical Hamiltonian equations have the form
\begin{subequations}
\label{EQ15}
\begin{eqnarray}
\dot{z} &=& -2\frac{\pd H}{\pd \Phi}=\frac{\Lambda}{2}(1-z^2)\sin\Phi,
\label{EQ15a}\\
\dot{\Phi} &=& 2\frac{\pd H}{\pd z}= (1-2\Lambda-\Lambda\cos\Phi)z.\label{EQ15b}
\end{eqnarray}
\end{subequations}
In our case these equations correspond to the mean field approximation.

In order to clarify the physical meaning of the introduced classical variables $z$
and $\Phi$ let us compute
\begin{align}
\langle \Psi|(n_2-n_1)|\Psi\rangle=N\sum_{N_1,N_2\atop N_1+N_2=N}
C_{N_1}^*\frac{N_2-N_1}{N}C_{N_1}
\nonumber
\\
=N\left(1-2\sum_{k=0}^N\frac{k}{N} |C_k|^2\right)=N(1-2\langle x\rangle),
\label{EQ16}
\end{align}
and
\begin{align}
 \langle\Psi|(b_2^\dag)^2b_1^2|\Psi\rangle=N^2\sum_{k=0}^{N}C_k^*b_{k+1}C_{k+2}
 \nonumber \\
 = N^2\langle b_h\left(x+\frac h2\right )e^{i\hat{p}}\rangle
     \label{EQ17}
\end{align}
(in the last formula we used that $C_{k+2}=e^{i\hat{p}}C_{k}$).

Thus in the limit $h\to 0$
\eqb
z=\lim_{h\to 0} \langle z\rangle = \frac{ \langle n_2\rangle-\langle n_1\rangle}{N}
\eqe
is the relative population of the states (as it is clear $z\in [-1,1]$)
and
\eqb
\Phi =\lim_{h\to 0} \langle \hat{p}\rangle=
\mathrm{arg}\langle(b_2^\dag)^2b_1^2\rangle.
\label{EQ18}
\eqe
is the relative phase.

\subsection{On the mean-field dynamics}
\label{sec:meanfield}

Let us now discuss some aspects of the dynamics described by the mean-field
approximation. First of all we observe that the system (\ref{EQ15})  has two fixed
points: $P_1= (z=0,\Phi=0)$ and $P_2 = (z=0,\Phi=\pi)$. The respective frequencies
are $\Omega^2_1 = \frac{3\Lambda}{2}(\Lambda-\Lambda_c)$ and $\Omega^2_2 =
\frac{\Lambda}{2}(1-\Lambda)$ with $\Lambda_c = 1/3$. Thus one easily verifies that
$P_2$ is a local maximum, while $P_1$ is a saddle-point for $\Lambda<\Lambda_c$ and
a local minimum for $\Lambda>\Lambda_c$.

Next, following Ref.~\cite{braun} we take into account that the  classical energy
$E={\cal H}(z,\Phi)$ is bounded by the two potential curves, one corresponding to
$\Phi=0$ [the curve $U^{(+)}(x)$] and the other to $\Phi=\pi$ [the curve
$U^{(-)}(x)]$:
\eqb
U^{(-)}(x) = (\Lambda-1)b_0(x),\; U^{(+)}(x)=(3\Lambda-1)b_0(x),
\label{potcurves}\eqe
i.e. $U^{(-)}(x)\le E \le U^{(+)}(x)$.  Therefore, the turning points of the
classical system (\ref{EQ15a})-(\ref{EQ15b}) lie on the curves $U^{(\pm)}(x)$. One
can express the period of oscillations between two turning points $x_1$ and $x_2$
as follows
\eqb
T = 2\int\limits_{x_1}^{x_2}\frac{\rd x}{\sqrt{(U^{(+)}-E)(E-U^{(-)})}}.
\label{period}
\eqe
This integral can be expressed in terms of the complete elliptic  integral of the
first kind $K(\cdot)$. To this end we single  out four physically different cases:

{\em Case 1}: If $\Lambda <\Lambda_c$ and the energy of ``classical"  motion is
$(3\Lambda-1)/4<E<0$, the turning points belong to different curves, for instance:
\begin{eqnarray*}
&&x_1=\frac 12\left( 1-\sqrt{1+\frac{4E}{1-\Lambda}}\right)\in U^{(-)}
\\
&&x_2=\frac 12\left( 1-\sqrt{1+\frac{4E}{1-3\Lambda}}\right)\in U^{(+)}
\end{eqnarray*}
 and the period is given by
\begin{eqnarray}
    \label{period_1}
    \displaystyle{
    T_1= \frac{4  K\left(\frac{2\sqrt{-2E\Lambda}}{\sqrt{(1-3\Lambda)(1-\Lambda+4E)}}\right)}
    {\sqrt{(1-3\Lambda)(1-\Lambda+4E)}}\,.
    }
\end{eqnarray}
The difference in the atomic population oscillates about a nonzero value.

{\em Case 2}: If $\Lambda <\Lambda_c$ and the energy of ``classical"  motion is
$(\Lambda-1)/4<E<(3\Lambda-1)/4<0$, the turning points belong to a single curve:
\begin{eqnarray*}
x_{1,2}=\frac 12\left( 1\pm\sqrt{1+\frac{4E}{1-\Lambda}}\right)\in U^{(-)}
\end{eqnarray*}
 and the period is given by
\begin{eqnarray}
    \label{period_2}
    \displaystyle{
    T_2= \frac{2\sqrt{2}}{\sqrt{-\Lambda E}}
    K\left(\frac{\sqrt{(1-3\Lambda)(1-\Lambda+4E)}}{2\sqrt{-2E\Lambda}}\right)\,.
    }
\end{eqnarray}
The difference in the atomic population oscillates about zero.

{\em Case 3}: If $\Lambda >\Lambda_c$ and the energy of ``classical"  motion is in
the interval $(\Lambda-1)/4<E<0$, the turning points belong to a single curve:
\begin{eqnarray*}
x_{1,2}=\frac 12\left( 1\pm\sqrt{1+\frac{4E}{1-\Lambda}}\right)\in U^{(-)}
\end{eqnarray*}
 and the period is given by
\begin{eqnarray}
    \label{period_3}
    \displaystyle{
    T_3= \frac{8K\left(\sqrt{\frac{1+\frac{4E}{1-\Lambda}}{ 1+ \frac{4E}{1-3\Lambda}}}\right) }
    {\sqrt{(1-\Lambda)(3\Lambda-1-4E)}}
     \,.
    }
\end{eqnarray}
The two states have equal average populations.

{\em Case 4}: If $\Lambda >\Lambda_c$ and the energy of ``classical" motion is in
the interval $0<E<(3\Lambda-1)/4$, the turning points belong to a single curve:
\begin{eqnarray*}
x_{1,2}=\frac 12\left( 1\pm\sqrt{1+\frac{4E}{1-3\Lambda}}\right)\in U^{(+)}
\end{eqnarray*}
 and the period is given by
 \begin{eqnarray}
    \label{period_4}
    \displaystyle{
    T_4= \frac{8K\left(\sqrt{\frac{1+\frac{4E}{1-3\Lambda}}{ 1+\frac{4E}{1-\Lambda}}}\right)}
    {\sqrt{(3\Lambda-1)(1-\Lambda-4E)}}
    \,.
    }
\end{eqnarray}
The two states have equal average populations.

Finally we mention that equations (\ref{EQ15}) can be directly obtained  from the
two-mode Hamiltonian of the Bloch-band tunneling in the meanfield approximation
(see \cite{BKKS,comm1} for more details):
\begin{eqnarray}
\label{ham2}
    H_{m-f}=  \chi_{11}|{\cal A}_1|^4+  \chi_{11}|{\cal A}_2|^4 +4\chi_{12}|{\cal A}_1|^2|{\cal A}_2|^2
    \nonumber \\
    +  \chi_{12} ({\cal A}_1^*)^2{\cal A}_2^2+  \chi_{12}({\cal A}_2^*)^2{\cal A}_1^2\,.
\end{eqnarray}
where the complex amplitudes ${\cal A}_{1,2}$ are determined by the expansion of the order parameter $\Psi$
\begin{eqnarray}
\label{psi}
\Psi={\cal A}_{1}(t)\varphi_{1}({\bf r})e^{-iE t}+ {\cal A}_{2}(t)\varphi_{2}({\bf
r})e^{-iE t}\,
\end{eqnarray}
and determine average populations of the levels ${\cal N}_j=|{\cal A}_{j}|^2$,  and
$\varphi_j(\br)$ are the respective Bloch states. The complex amplitudes are
connected by the particle conservation law: $N={\cal N}_1+{\cal N}_2$.  Now the
system (\ref{EQ15a}), (\ref{EQ15b}) is obtained from the Hamiltonian equations by
defining
\begin{eqnarray}
\label{z_phi_mean}
z=\frac{|{\cal A}_2|^2-|{\cal A}_1|^2}{{\cal N}}\quad\mbox{and}\quad \Phi= \arg\{{\cal A}_1^2 ({\cal
A}_2^*)^2\}
\end{eqnarray}
[c.f. Eq. (\ref{EQ18})].

\subsection{``Coherent'' states}

Turning now to the quantum system, we observe that the   dynamical equations for
the coefficients $C_n$ have to be supplied by the initial conditions. As in the
standard WKB approximation, the corresponding initial conditions must be smooth
enough. At the same time a natural question arises about construction of quantum
states, $|\Psi(t)\rangle_c$, most closely resembling the meanfield dynamics. We
will refer to such states as coherent states.

In order to construct such states we recall the explicit form of the  meanfield
ansatz (\ref{psi}) and consider the respective boson operator $c^\dag = \alpha_1
b^\dag_1 +\alpha_2 b^\dag_2$, where $\alpha_{1,2}$ are time dependent complex
parameters satisfying $|\alpha_1|^2 +|\alpha_2|^2 = 1$. Next we define
\begin{eqnarray}
|\Psi(t)\rangle_{c} &\equiv& |\alpha_1,\alpha_2\rangle \equiv\frac{(\alpha_1(t) b^\dag_1 +\alpha_2(\textbf{}t)
b^\dag_2)^N}{\sqrt{N!}}|0,0\rangle
\nonumber \\
 &=&\sum_{k=0}^N\sqrt{\frac{N!}{k!(N-k)!} }\alpha_1^k\alpha_2^{N-k}|k,N-k\rangle.
\label{EQ19}
\end{eqnarray}

By using the identities:
\begin{eqnarray*}
&&\langle\alpha_1,\alpha_2 |n_j|\alpha_1,\alpha_2 \rangle = N|\alpha_j|^2,
\\
&&\langle\alpha_1,\alpha_2 |n_j^2|\alpha_1,\alpha_2 \rangle = N|\alpha_j|^2 + N(N-1)|\alpha_j|^4,
\\
&&\langle\alpha_1,\alpha_2 |n_1n_2|\alpha_1,\alpha_2 \rangle = N(N-1)|\alpha_1\alpha_2|^2,
\\
&&\langle\alpha_1,\alpha_2 |(b_2^\dag)^2b_1^2|\alpha_1,\alpha_2\rangle = N(N-1)\alpha_1^2(\alpha_2^*)^2,
\end{eqnarray*}
we obtain for the energy (dropping an inessential constant)
\begin{eqnarray}
& & E = \langle\alpha|H|\alpha\rangle = \chi_{11}N(N-1) \nonumber\\
& & \times\left\{|\alpha_1|^4 +|\alpha_2|^4 +\Lambda[4|\alpha_1|^2|\alpha_2|^2
+\alpha_1^2(\alpha_2^*)^2 + \alpha_2^2(\alpha_1^*)^2]\right\}.\nonumber\\
&& \label{EQ20}\end{eqnarray}
Writing  the Hamilton equations for the complex amplitudes $\alpha_j$ in the form
$$
i\hbar\frac{\rd\alpha_j}{\rd t} = \frac{1}{ N}\frac{\pd E}{\pd \alpha_j^*},
$$ in
terms of the normalized time $\tau = 4\chi_{11}N/\hbar$ and setting
\begin{eqnarray}
\label{z_phi_coher}
z =
|\alpha_2|^2 - |\alpha_1|^2\quad\mbox{and}\quad \Phi = \arg\{\alpha_1^2\alpha_2^{*2}\}
\end{eqnarray}
[c.f. Eq. (\ref{z_phi_mean})] we recover the system (\ref{EQ15}) in the limit
$N\to\infty$. Comparing  (\ref{ham2}) with (\ref{EQ20})  one gets the physical
meaning and a link among the quantum mechanical and mean-field amplitudes
$\alpha_j$ and ${\cal A}_j$.

\section{Numerical simulations}
\label{numer}

In order to proceed with the numerical simulation of the described phenomena we
observe that their occurrence does not depend on a particular choice of the
potential. Therefore we concentrate on the simplest case of a lattice
having cos-like profile. Recalling that the spatial coordinates are measured in the
units of the lattice period $d$, the energy is measured in terms of the recoil
energy $E_r = \hbar^2\pi^2/(2md^2)$, the time is measured in the units $\hbar/E_r$
and the lattice is $\pi$-periodic, we set
\eqb
V = V_0[\cos(2x) + \cos(2y)].
\label{LATT}
\eqe
The respective BZ is given by $[-1,1]\times[-1,1]$.

We consider intra-band tunneling between two  X-points of the second lowest band.
The respective tensors of the inverse effective mass have positive components, and
therefore the respective Bloch states are modulationally stable, provided the
interactions among atoms are repulsive (i.e. $\chi_{ij}>0$).

Generally speaking one may have two physically different situations. When $V_0>
V_{\mathrm{th}}$, with $V_{\mathrm{th}}\approx 0.627$ for the case (\ref{LATT}),
there exists a full two-dimensional  gap in the spectrum, whose  width depends on a
particular value of $V_0$. If the potential amplitude is weak enough, i.e. $V_0 <
V_{\mathrm{th}}$, the gap is closed (the gap width becomes zero at $V_0 =
V_{\mathrm{th}}$). For $V_0 = V_{\mathrm{th}}$ one computes, using (\ref{LATT}),
that $\chi^{22}_{\bq_1\bq_1} \approx 0.1494$ and $\chi^{22}_{\bq_1\bq_2} \approx
0.1303$, and respectively $\Lambda= \Lambda_{\mathrm{th}} \approx 0.8727
>\Lambda_c$. Thus, generally speaking, one can distinguish three different regions
of the parameters, which correspond to (i) $\Lambda < \Lambda_c$, (ii) $\Lambda_c <
\Lambda < \Lambda_{\mathrm{th}}$, and (iii) $\Lambda> \Lambda_{\mathrm{th}}$. The
parameters $\Lambda_{\mathrm{th}}$ and $\Lambda_c$ have however different physical
origins: $\Lambda_{\mathrm{th}}$ is related to the band structure and thus can
affect physical applicability of the two mode model, due to possible tunneling to
the other bands, while $\Lambda_c$ is an intrinsic parameter of the model.

Since our numerical study aims to check precisely the two-mode  model we will
consider only the case where the full gap is open, i.e.  $V_0>V_{\mathrm{th}}$, and
select two particular cases: the Case I with $V_0=6$ when
$\chi_{\bq_1\bq_1}^{22}\approx 0.2708$ and $\chi_{\bq_1\bq_2}^{22}\approx
0.0.0652$, and respectively $\Lambda\approx 0.2407<\Lambda_c$, and the Case II with
$V_0=1$ when $\chi_{\bq_1\bq_1}^{22}\approx 0.1551$ and
$\chi_{\bq_1\bq_2}^{22}\approx 0.0626$, and respectively $\Lambda\approx
0.4040>\Lambda_{c}$.

The numerical simulations of the discrete Schr\"odinger equation (\ref{EQ9}) were
performed  by using the variable order Adams-Bashforth-Moulton  solver. The error
was controlled by checking the norm of the vector $C_k(t)$. The values of $z$ and
$\Phi$ were obtained by using the correspondence formulas (\ref{EQ16}) and
(\ref{EQ17}).

\begin{figure}[ht]
\begin{center}
\includegraphics[scale=0.5, bb = 53   201   517   579 ]{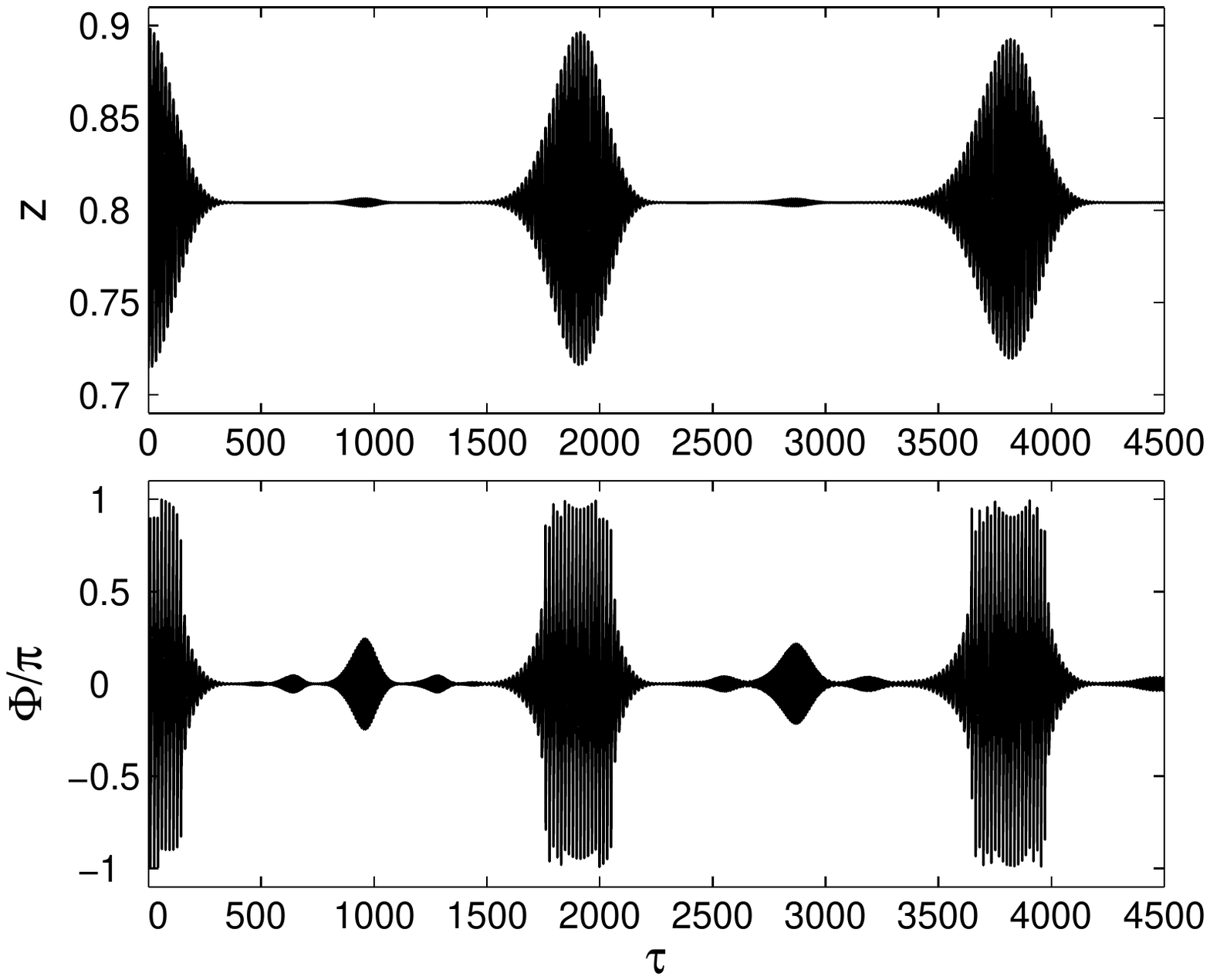} \caption{Quantum
evolution for $N=350$ BEC atoms and the lattice with $\Lambda=0.2407$. As the
initial population we used the Gaussian  distribution $C_k(0) = \exp\{ik -
(k-k_c)^2/(2\sigma^2)\}$ with  $\sigma^2 = 50$ and $k_c=50$.}
\label{FG2}
\end{center}
\end{figure}

In Fig.~\ref{FG2} we show typical dynamics of the populations in the  Case I, i.e.
for $\Lambda<\Lambda_c$, of the condensate of $N=350$ atoms. One can observe
relatively fast, i.e. at $\tau\sim 100$, decay of Rabi oscillations followed by
almost steady distribution of the atoms, approximately 90\% of atoms concentrated
in one state. This is behavior is typical for a quantum collapse. The both
populated states are characterized by the same phases. After much longer interval
of time $\Delta \tau \sim 1500$ the revival of the oscillations is observed.

\begin{figure}[ht]
\begin{center}
\includegraphics[scale=0.5, bb = 53   201   513   578 ]{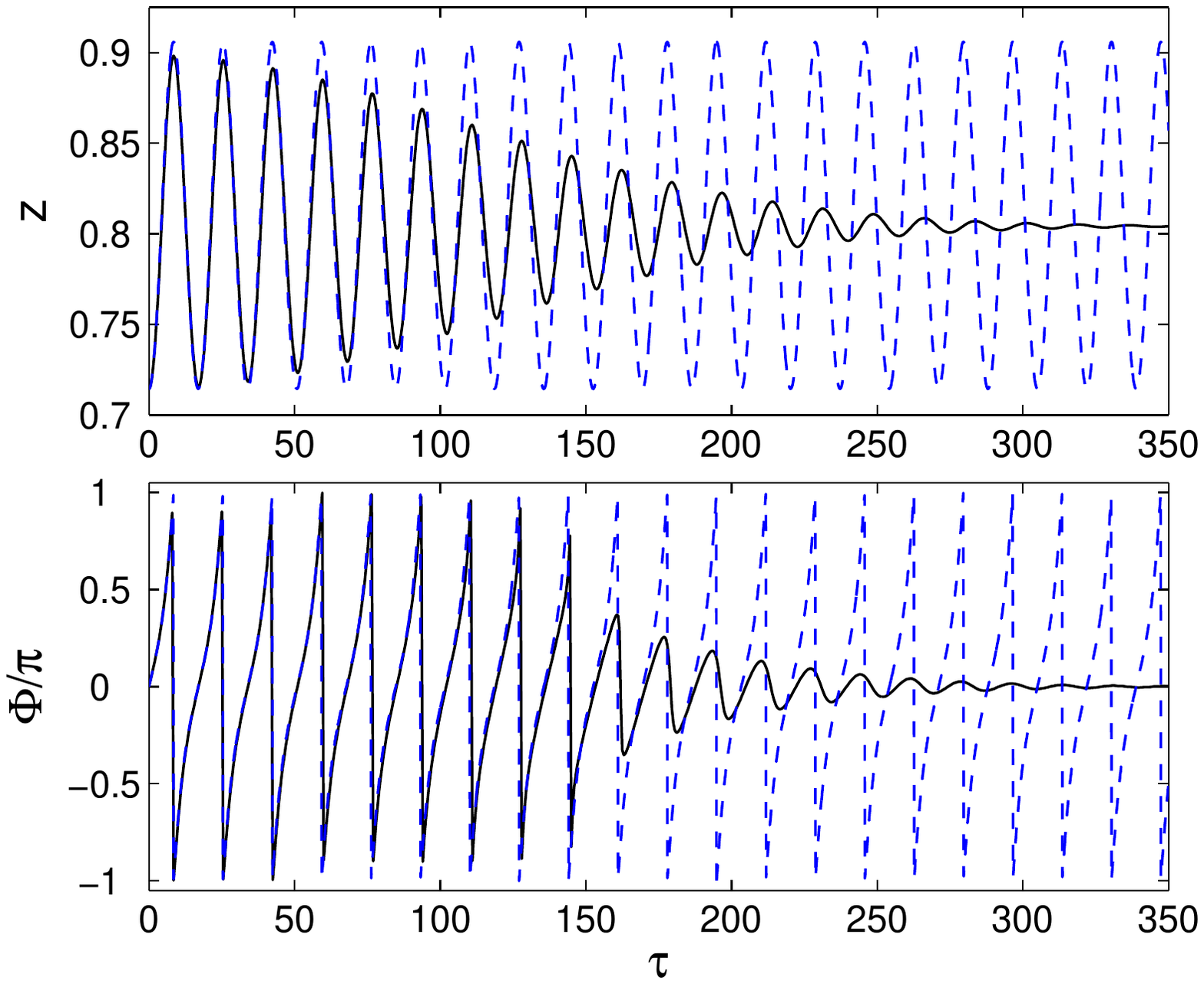}
\caption{\label{FG3} The initial quantum evolution (solid lines) vs meanfield
evolution (dashed lines) for parameters as in Fig. \ref{FG2}.}
\end{center}
\end{figure}

Although collapse of Rabi oscillations was earlier observed in  pure mean-field
models (see ~\cite{KKS,BKK}), the nature of the phenomenon considered here is very
different. Suppression of the oscillations in the case of spatially extended
systems is related to development of inhomogeneous spatial patterns, mainly related
to modulational instability of one of the states. In the case at hand, however the
effect is essentially quantum and disappears in the meanfield limit (where
$N\to\infty$). This is clearly illustrated in Fig.~\ref{FG3}, where we compare the
results of the quantum dynamics with its mean-field limit at earlier stages of the
evolution.

\begin{figure}[ht]
\begin{center}
\includegraphics[scale=0.5, bb = 53   201   521   586 ]{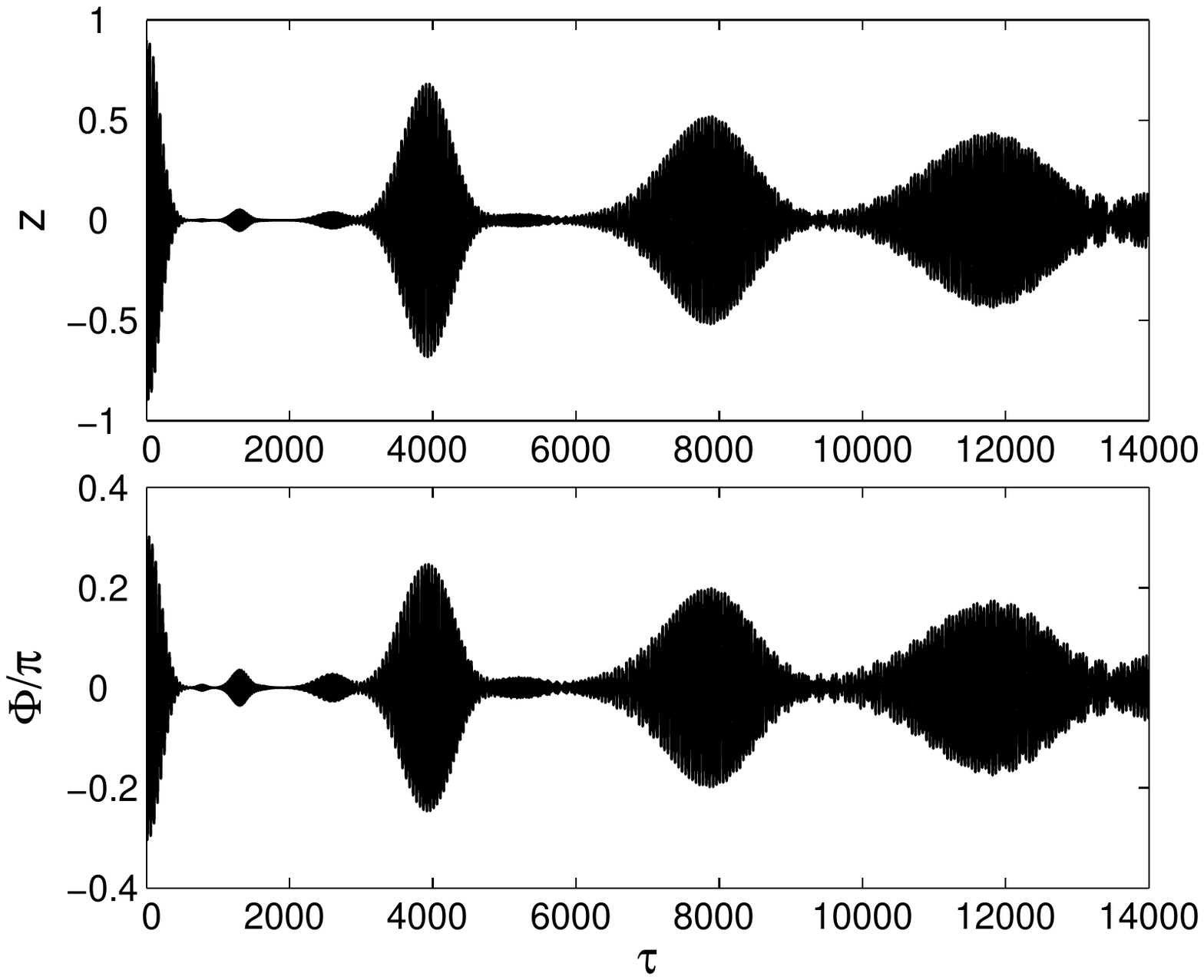} \caption{\label{FG4}
Quantum evolution for $N=500$ BEC atoms and the lattice with $\Lambda=0.404$. The
initial condition used is $C_k(0) = \exp\{- (k-k_c)^2/(2\sigma^2)\}$ with $k_c= 25$
and  $\sigma^2 = 50$.}
\end{center}
\end{figure}

\begin{figure}[ht]
\begin{center}
\includegraphics[scale=0.5, bb = 53   201   521   586 ]{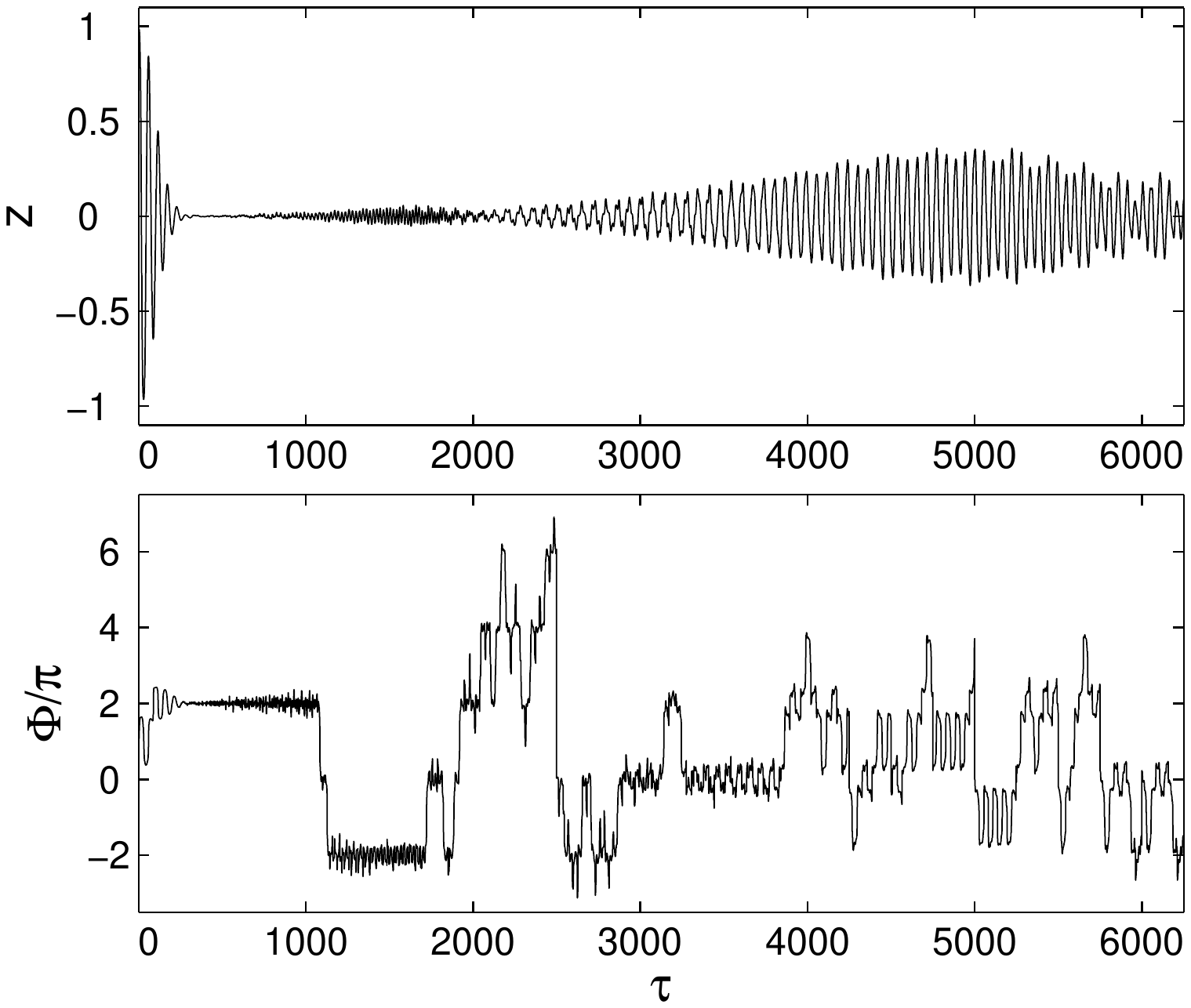}
\caption{\label{FG5} Quantum evolution for $N=1000$ BEC atoms and the lattice with
$\Lambda=0.404$.  The initial condition used is $C_k(0) = \exp\{ik -
(k-k_c)^2/(2\sigma^2)\}$ ($\Phi(0) = 2$) with $k_c = 10$ and $\sigma^2 = 50$.}
\end{center}
\end{figure}

Now we turn to the Case II, where $\Lambda>\Lambda_c$, which is shown  in
Figs.~\ref{FG4} and \ref{FG5}. The main feature of this situation is that as a
result of quantum collapse both states become equally  populated ($z\to 0$), what
is also in accordance with the meanfield dynamics (c.f. Figs.~\ref{FG2} and
\ref{FG4}). The revival occurs at latter times $\Delta\tau\sim 3000$.

In all the figures one can observe that the  period of the Rabi oscillations is
very accurately reproduced by the mean-filed approximation, i.e. by the formulas
(\ref{period_1}), (\ref{period_2}), (\ref{period_3}), and (\ref{period_3}). In
particular, Figs.~\ref{FG3} (or ~\ref{FG2}) and ~\ref{FG5} correspond to the Cases
1 ( with $z=0.72$ and $E\approx -0.033$) and 3 (with $z=0.98$ and $E\approx
-0.0059$) described in Sec.~\ref{sec:meanfield}. Then using (\ref{period_1}) and
(\ref{period_4}) one computes $T\approx 16.86$ and $T\approx 51.99$, what matches
very well with the periods obtained from the direct numerical simulations. The
described behavior has simple physical explanation: passage to the mean-field
description is performed at a constant $gN$, i.e. at a constant effective
nonlinearity of the system. Meantime, the tunneling time depends on the relative
value of the nonlinear interaction term, i.e. on $\Lambda$.

Comparing Figs.~\ref{FG2} and~\ref{FG4} one  observes the expected effect of the
delay of the quantum collapse in terms of the dimensionless time $\tau$ with
increase of the number of particles. At the same time the  numerical simulations
did not reveal any significant effect of the initial phase mismatch $\Phi(0)$ on
the collective dynamics.

Another relevant parameter is the initial distribution of the atoms. Clearly, the
closer is the initial distribution of atoms (defined  by $k_c$ and $\sigma$ in the
numerics) to the one violating the mean-filed assumptions, when almost all of the
atoms are at one of the X${}_j$ points, the further is the dynamics from the
quasi-classical one, which is reflected in Fig. \ref{FG5} in the fact that the
collapse time of the oscillations is twice as smaller as that of Fig. \ref{FG4} for
twice as much number of atoms. Also, in Fig. \ref{FG5} the phase dynamics at the
recurrence of quasi-classical relative population oscillations contains the
quasi-classical oscillations interrupted  by the $2\pi m$ jumps.

\section{Discussion and conclusion}

In the present paper we have considered the quantum  tunneling of a Bose-Einstein
condensate loaded in a two dimensional lattice. The considered tunneling occurs
between two stable states, hence allows us to restrict the consideration to a
spatially homogeneous model. Such a model, in the rotating wave approximation, was
further reduced to the effective two-state system, where both states possess the
same energies. The considered tunneling  is essentially nonlinear and related to
the four wave-mixing due to the two-body interactions (in a pure linear system such
a tunneling cannot occur because of violation of the momentum conservation).

The main finding of the work is the quantum  collapse and revivals of the Rabi
oscillations between the tunneling states, which are suppressed  in the mean-field
approximation due to the fact that tunneling between two equal-energy states result
in configurations of low atomic population of one of the states and thus, strictly
speaking, cannot be described in terms of the  mean-field approximation.
Nevertheless, the latter approach turns out to be useful in predicting  different
regimes of the tunneling, with collapse occurring to either equal or disbalanced
populations of the states, and for accurate estimate of the frequency of the Rabi
oscillations (i.e. the tunneling time).

Existence of a  well-defined frequency of the oscillations and the difference in
spatial patterns of the condensate in different X-states of the lattice suggest a
way of  experimental observation of the phenomenon, which could be based, for
example, on direct imaging at specific moments of time. To estimate the physical
time scale corresponding to the period of the Rabi oscillations, we consider a
condensate of $^{87}$Rb atoms, with the $s$-wave scattering length $a_s=5.25 nm$,
which is tightly trapped, say in $y$ direction, with the respective linear
oscillator length $\ell = 0.1\,\mu$m, and has $N=350$ atoms occupying $M=25$ sites
of a lattice with the period $d=2\,\mu$m (i.e. the 2D optical lattice is imposed in
the $(x,z)$ plane and has 5 cells in each direction). In this case $g$ (the 2D
nonlinear constant) is given by $g=2\sqrt{2\pi}\hbar^2 a_s/m\ell_y$. Then using the
data from Fig.~\ref{FG2} (also Fig.~\ref{FG3}) we obtain that the dimensionless
period $T\approx 16.68$ in the physical units is $T_{Rb}\approx0.00655\,$s. Thus,
taking into account that the collapse and subsequent revival would occur after
about $1\,$s and $10\,$s, respectively, we see that this is a relatively slow
process. This is an expected slowness as far as in this example we used a small
number of atoms not providing effective enough tunneling due to the two-body
interactions. The collapse and revival time could be reduced by using a larger
condensate, however, for a very large number the collapses and revivals  will be
suppressed because of approaching the mean-field limit. An alternative way to
accelerate the quantum collapse, and thus make easier its observation is to use
more light atoms, say sodium ones, or by increasing the scattering length.

A number of important questions, however, are left for further studies. Among them
we mention the resonant interaction of more than two states, the quantum theory of
nonlinear Landau-Zener tunneling, the study of interplay of quantum collapse and of
modulational (dynamical) instability in the case when the spatial extension of the
system is taken into account and the states among which the tunneling occurs
possess different stability properties (what is an intrinsic feature of the
nonlinear systems), the effect of opening or closer of the total gap in the
lattice, the interplay between the quantum collapse and the dynamical collapse
(also called the blow up) which can occur in Bose-Einstein condensates with
attractive atomic interactions, etc.

\acknowledgements We are grateful to A. M. Kamchatnov for invaluable discussions at
the initial stage of this work. V.V.K. is grateful to V. M. P\'erez-Garc\'{\i}a for
the warm hospitality at the Department of Mathematics of the Universidad de
Castilla-La Mancha. The work of V.V.K. was supported by the Secretaria de Stado de
Universidades e Investigación (Spain) under the grant SAB2005-0195 and by the FCT
and European program FEDER under the grant POCI/FIS/56237/2004. The work of V.S.S.
was supported by the research grants from the CNPq and CAPES of Brazil.

\end{document}